\documentclass[%
 reprint,
 amsmath,amssymb,
 aps,
]{revtex4-2}

\usepackage{graphicx}
\usepackage{dcolumn}
\usepackage{bm}

\usepackage{algorithm}
\usepackage{algpseudocode}
\usepackage{xcolor}

\begin{document}


\title{Optimal ambition in business, politics and life}

\author{Ekaterina Landgren}
\affiliation{Department of Environmental Social Sciences, \\ Stanford Doerr School of Sustainability, \\ Stanford University, Stanford, CA 94305, USA}
\affiliation{
Cooperative Institute for Research in Environmental Sciences,\\ University of Colorado Boulder, Boulder, CO 80309, USA
}
\email{kath.landgren@stanford.edu}

\author{Ryan E. Langendorf}
\affiliation{
Department of Environmental Studies,\\ University of Colorado Boulder, \\ Boulder, CO 80303, USA
}%

\author{Matthew G. Burgess}
\affiliation{%
Department of Economics, College of Business, \\ University of Wyoming, Laramie, WY 82071, USA
}%

\date{\today}

\begin{abstract}
In business, politics and life, folk wisdom encourages people to aim for above-average results, but to not let the perfect be the enemy of the good. Here, we mathematically formalize and extend this folk wisdom. We model a time-limited search for strategies having uncertain rewards. At each time step, the searcher either is satisfied with their current reward or continues searching. We prove that the optimal satisfaction threshold is both finite and strictly larger than the mean of available rewards---matching the folk wisdom. This result is robust to search costs, unless they are high enough to prohibit all search. We show that being too ambitious has a higher expected cost than being too cautious. We show that the optimal satisfaction threshold increases if the search time is longer, or if the reward distribution is rugged (i.e., has low autocorrelation) or left-skewed. The skewness result reveals counterintuitive contrasts between optimal ambition and optimal risk taking. We show that using upward social comparison to assess the reward landscape substantially harms expected performance. We show how these insights can be applied qualitatively to real-world settings, using examples from entrepreneurship, economic policy, political campaigns, online dating and college admissions. We discuss implications of several possible extensions of our model, including intelligent search, reward landscape uncertainty and risk aversion.
\end{abstract}

\maketitle


\section{\label{sec:introduction} Introduction}

Norman Vincent Peale famously said, ``Shoot for the Moon. Even if you miss, you'll land among the stars." However, folk wisdom suggests that it is actually possible to aim too high, just as it is possible to aim too low. In a business context, entrepreneurs are encouraged to pursue big ideas, but not unrealistic pie-in-the-sky ones. Career advisers encourage people to build skills and social capital gradually, rather than expecting their first job to be perfect (e.g., ref. \cite{jay2012defining}), and to be opportunistically mobile but not restless or noncommittal. Common dating advice encourages people to have realistic expectations about what to look for in a partner. In politics, incrementalism encourages gradual changes to improve society without destabilizing its institutions \cite{lindblom1959science}.  

Empirical research seems to support these folk intuitions encouraging intermediate ambition. For example, intermediate procrastinators are often the most successful entrepreneurs and innovators \cite{grant2017originals}. They do not over-commit to their first ideas, but they still have the focus to execute better ideas later. Employees' propensities to change jobs tend to have mixed effects on career success and professional well-being \cite{guan2019career}. Research supports incrementalism as a policymaking strategy, due to checks and balances, distributed expertise, and pluralistic social and political interests \cite{lindblom1959science,hayes2002limits}.  The happiest marriages, on average, occur between people who are similar to each other in traits related to desirability (e.g., health, vitality, exercise, spirituality) \cite{george2015couple}. Natural resource exploiters such as fishers, who do not have access to perfect information about where the resources are most plentiful, maximize their catch by re-visiting patches of slightly above-average quality \cite{burgess2020opportunities}. These are a few examples of many.

Here, we mathematically formalize and extend these folk intuitions about ambition. We show in a general search model that optimal ambition targets outcomes that are both strictly finite and strictly larger than the mean of available rewards. In other words, optimal ambition \textit{does not} `shoot for the moon', but it \textit{does} try to do better than average. Our model has conceptual parallels to areas of applied mathematics, economics and finance, and theoretical biology related to optimal search and optimal stopping, which we discuss. 

We also show how optimal ambition depends on certain features of the reward landscape, and how optimal ambition is affected by social comparison and search costs. We discuss other possible extensions of our model. 

Our results provide a precise and accessible conceptual link between the diverse domains in which intermediate ambition seems to be supported. Our model also provides qualitative insights for optimally calibrating ambition, which should generalize beyond our simplifying assumptions. We illustrate some of these insights using empirical examples from economic policy, entrepreneurship and political campaigns. Our model provides testable hypotheses for empirical research, which we illustrate using examples from online dating and college admissions.

\section{Model}

We model an `agent' searching among a set of possible strategies over a finite number of time periods, $t_{\text{max}}$. The agent does not know each strategy's reward in advance, but they do have information about the statistical distribution of rewards across all strategies. We explore scenarios where the agents know the true distribution of rewards, and scenarios where the agents estimate the distribution from their peers' rewards.

Strategies and rewards could abstractly represent, for example, jobs and salaries for a job-seeker, colleges and application cycles for a prospective student, venture ideas and payoffs for an entrepreneur, returns and exercising time for a financial option holder, dating prospects and relationship satisfaction for a single, campaign strategies and popularity for a political candidate, policies and their outcomes for a policy maker, or patches with different potential harvests for a harvester. There are many other possible examples. Our model is admittedly simpler than these real decision settings, but we argue that its key qualitative insights should generalize to more complex settings. 

During each time period, our agent either sticks with their current strategy or chooses a new one. Their objective is to maximize the sum of rewards across the $t_{\text{max}}$ time periods. Their key decision variable is when to be satisfied with the current strategy. We assume that agents have a satisfaction threshold, $T$, measured as a number of standard deviations, $\sigma$, above or below the mean reward across all strategies, $\mu$. The agent searches in each time step until they find a strategy with a reward greater than or equal to $T$. Then, they stick with that strategy and its reward for the rest of the time periods. The higher $T$, the more ambitious the agent. 

The agent must balance the costs of being too easily satisfied and settling for less than what is attainable, with the costs of being too ambitious and passing on high, attainable rewards while searching fruitlessly for the unattainable. We prove that the optimal satisfaction threshold, $T$, is strictly larger than the mean, $\mu$ (i.e., agents should try to do better than average), but the optimal $T$ is also finite (i.e., overambition is possible and costly). We explore how the optimal satisfaction threshold depends on several properties of the search, the reward distribution, the information agents use to make their decisions and the search cost. 

\begin{figure}[t]
\centering
\includegraphics[width=\linewidth]{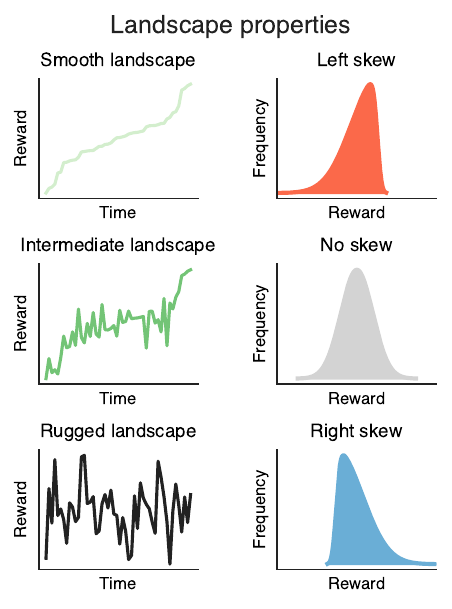}
\caption{\textbf{Reward landscapes.} Stylized representations of reward landscape ruggedness (left) and skewness (right). Ruggedness determines how different successive rewards can be from each other. Skewness determines the relative abundance of low and high rewards, among landscapes that share the same mean and variance.}
\label{fig:conceptual_figure}
\end{figure}

We analyze the smoothness (vs. ruggedness) and skewness of the reward distribution (Fig.~\ref{fig:conceptual_figure}). Smoothness describes the autocorrelation of successive rewards, measured with a parameter, $\varphi$. When $\varphi=0$, successive rewards are uncorrelated, and the reward landscape is maximally rugged. When $\varphi$ is close to 1, rewards are highly autocorrelated and the landscape is smooth. Fig.~\ref{fig:conceptual_figure} (left) illustrates smooth and rugged landscapes conceptually.

Left-skewed reward distributions have larger negative extremes than positive extremes, compared to the mean.  Right-skewed distributions are opposite. Fig.~\ref{fig:conceptual_figure} (right) shows skew-normal reward distributions, each having the same mean and standard deviation, but differing in their skewness.

With no search cost (we consider costs in section \ref{sec:search-costs} below), the agent's reward in time period $t$, $X_t$, is given by:

\begin{equation}
X_t=
\begin{cases}
\varphi X_{t-1}+(1-\varphi)\epsilon_t & X_{t-1}<T,\\ 
X_{t-1}  & X_{t-1} \geq T.
\end{cases}
\label{eq:1}
\end{equation}

Here, $\epsilon_t$ are independent and identically distributed (i.i.d.) samples from the reward distribution. Our model is effectively an autoregressive (AR(1)) random walk, which ends at time $t_{\text{max}}$ ($t\leq t_{\text{max}}<\infty$). This approach has the advantage of model parsimony while capturing the properties of commonly used fitness landscapes (e.g., the underlying distribution \cite{skellett2005maximally} and parent\textendash offspring correlation \cite{verel2003bottlenecks} in fitness landscapes, and autocorrelation behavior on general neighborhood digraph landscapes \cite{dimova2005arbitrary}). Equation \eqref{eq:1} describes the two phases of the search. In the ``explore'' phase, each agent conducts a random walk. Once the agent meets their respective threshold, they initiate the ``exploit'' phase and remain at the same reward value until they run out of time.

Our model offers a general, tractable, and accessible formulation applicable to a widely studied class of sequential decision problems that aim to maximize rewards (or minimize costs) under uncertainty (e.g.~\cite{macqueen1960optimal, van1974optimal, freeman1983secretary, powell2019unified, kohn1974theory}). Optimal stopping problems, concerned with determining the best time to take a particular action, have been widely applied in management science~\cite{ciocan2022interpretable} and finance~\cite{liu2022optimal}. Optimal search has been explored in economics \cite{kohn1974theory}, and in theoretical biology via optimal foraging theory \cite{pyke1984optimal}. For example, the Marginal Value Theorem~\cite{charnov1976optimal} has been used to explain how long diving mammals spend underwater~\cite{kramer1988behavioral}, how long birds spend in any given tree~\cite{cowie1977optimal} and how long bumblebees spend on flowers~\cite{schmid1985honeybees}. These same ideas have helped to explain the evolution of adaptations~\cite{parker1990optimality}. Depending on the context, solving such optimal stopping or optimal search problems may require advanced computational methods~\cite{becker2019deep, ciocan2022interpretable}. In contrast, in this work, we focus on a simple, self-contained formulation that emphasizes parsimony. For a detailed mathematical treatment of optimal stopping and hitting times in autoregressive processes, see refs. ~\cite{novikov2009distributions, christensen2012phase, christensen2011elementary}. We extend our model to incorporate features that are psychologically relevant and commonly reflected in human decision-making, such as varying reward distribution properties~\cite{guan2014threshold}, social comparison~\cite{guan2014threshold, gerber2018social, muller2010being} and search costs \cite{kohn1974theory}.

\section{Results}
\subsection{Expected rewards vs. satisfaction threshold}

\begin{figure}[!htb]
\centering
\includegraphics[width=\linewidth]{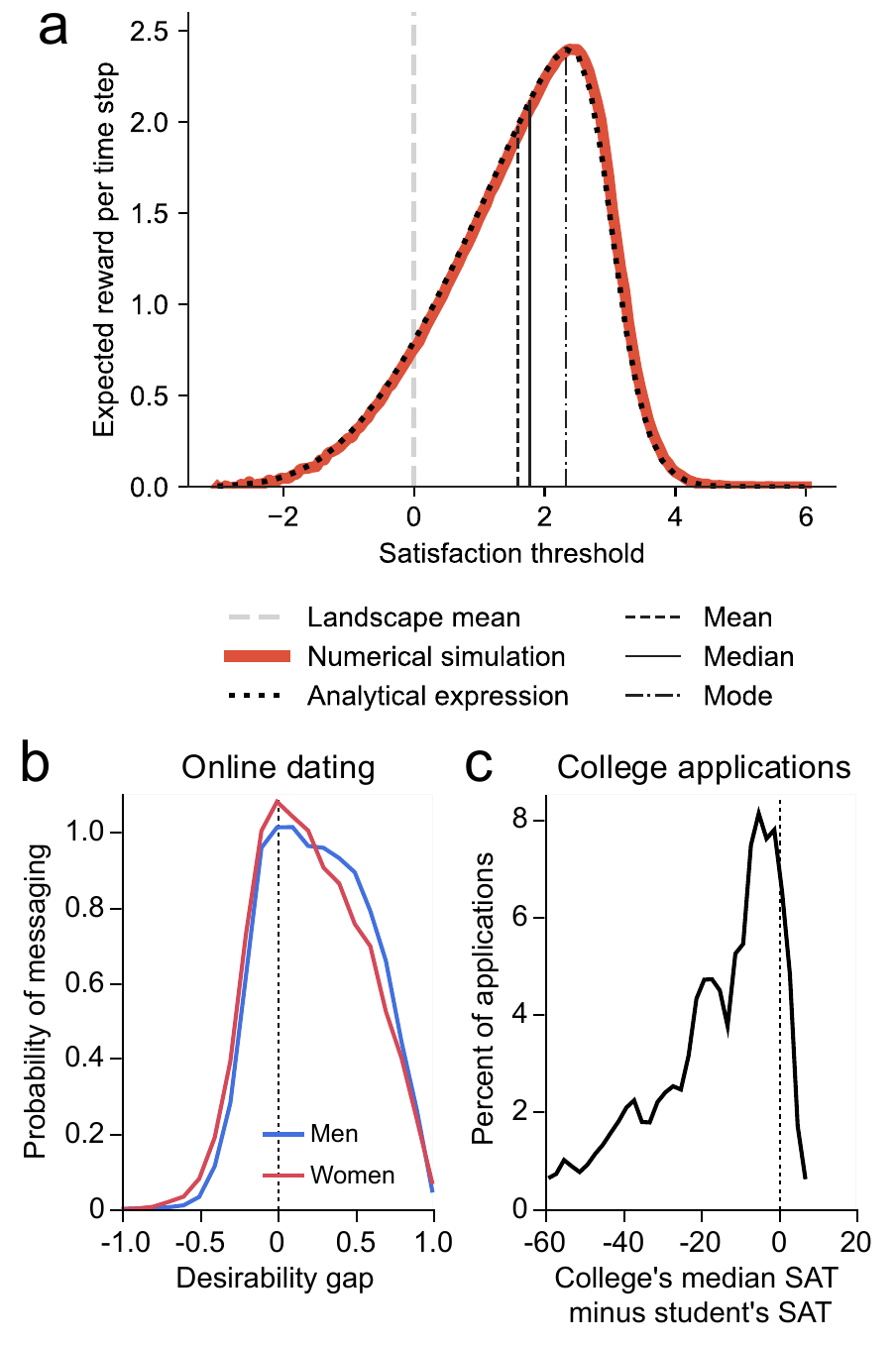}
\caption{\textbf{Expected rewards vs. satisfaction thresholds, and empirical examples.} Panel a: analytical and numerical calculations of the expected reward distribution, as a function of the target threshold, on a maximally rugged landscape. The expected reward distribution is unimodal and negatively skewed, with an optimal threshold above zero (the landscape mean). The analytical expression given by equation \eqref{eq:expected-reward} (black, dashed) matches the simulation results (red). Parameter values are: $\epsilon_t \sim \mathcal{N}(0,1)$, $\varphi=0$, and $t_{\text{max}}=1000$. Results shown are averaged over $10^4$ simulations. We note that many real-world applications occur on shorter time scales ($t_{\text{max}}\ll1000$), and consequently have lower optimal thresholds (e.g., optimal $T\approx 1.6$ with $t_{\text{max}}=100$). Panels b and c: examples of real-world search strategies. Panel b: When online dating, heterosexual men (blue) and women (red) are most likely to message potential partners who are slightly more desirable than they are (averaged over the four cities shown in fig. 2 of ref. \cite{bruch2018aspirational}). Panel c: When applying for college, members of the 2008 U.S. high-school graduating class concentrated their applications on schools with median SAT scores similar to, or slightly below, their own score (data from ref. \cite{hoxby2012missing}). This suggests students were either being sub-optimally ambitious, or they faced other constraints (e.g., on income or geography), as ref. \cite{hoxby2012missing} suggests was the case.}
\label{fig:theory-vs-sim}
\end{figure}

We can analytically derive the relationship between an agent's satisfaction threshold and their expected reward, in our model, in the special case of a maximally rugged landscape ($\varphi=0$) and a Gaussian distribution of rewards, $\epsilon_t \sim \mathcal{N}(\mu=0,\sigma^2)$. In other cases, we can accurately compute the relationship numerically (Fig.~\ref{fig:theory-vs-sim}a).

With a rugged landscape and a Gaussian reward distribution, with probability density function (PDF) $\phi(x)$ and cumulative density function (CDF) $\Phi(x)$, we compute the expected reward by thinking of the satisfaction threshold as a divider between the `explore' phase, where an agent is searching for a new strategy at each time step, and the `exploit' phase, where the agent is sticking with their satisfactory strategy and its reward. The rewards collected during the `explore' phase are, by definition, less than the threshold value (otherwise, the agent would stop exploring), so that $X_{\text{explore}}<T$. The rewards collected during the `exploit' phase are greater than or equal to the threshold value: $X_{\text{exploit}} \geq T$. Rewards earned during the exploration phase have the lower-tail truncated normal distribution with the mean $\mu_{\text{explore}}$, and the single reward which is repeatedly collected during the exploitation phase has the upper-tail truncated normal distribution with the mean $\mu_{\text{exploit}}$. 

For threshold $T$, the mean $\mu_{\text{explore}}(T)=-\dfrac{\phi(T)}{\Phi(T)}$ is given by the inverse Mills ratio \cite{greene2003econometric}. The agents expect to do worse than the landscape mean in the exploration phase, on average ($\mu_{\text{explore}}<0$). The expected reward earned during the exploitation phase is given by $\mu_{\text{exploit}}(T)=\dfrac{\phi(T)}{1-\Phi(T)}>0$. 

Let $0\leq t_x \leq t_{\text{max}}$ be the number of time steps spent in the exploration phase before a success. The agent will draw $t_x$ samples during the exploration phase and $t_{\text{max}} - t_x$ samples during the exploitation phase. At each time step, the probability of successfully meeting the threshold is equal to $1-\Phi(T)$. Therefore, we can analytically express the expected cumulative reward for a given agent by summing over all the possible lengths of the exploration phase:

\begin{align}
\mathbb{E}[\text{reward}] &= \sum_{t_x=0}^{t_{\text{max}}} \Bigg[ \mu_{\text{explore}}(T) t_x + \mu_{\text{exploit}}(T) \left( t_{\text{max}} - t_x \right) \Bigg] \nonumber \\
&\quad \times (1 - \Phi(T)) \Phi(T)^{t_x}.
\label{eq:expected-reward}
\end{align}

To standardize our results across different time spans ($t_{\text{max}}$), we normalize the cumulative reward by dividing by $t_{\text{max}}$ and compare expected reward per time step (e.g., Fig.~\ref{fig:theory-vs-sim}a).

The expected reward has a unimodal (i.e., hump-shaped) relationship with the satisfaction threshold (Fig.~\ref{fig:theory-vs-sim}a). Being always satisfied or never satisfied yields an expected cumulative reward of $\mu t_{\text{max}}$. Always-satisfied agents perpetually receive the reward of the first strategy they sample---the mean, $\mu$, on average. Never-satisfied agents randomly sample the distribution in each time period, earning an expected average reward of $\mu$. There is an intermediate range of satisfaction thresholds that earn higher expected rewards. 

In Appendix~\ref{app:derivations}, we prove that the optimal threshold ($T$) is strictly greater than the mean reward ($\mu$). The proof is somewhat complex, but the intuition is simple. Suppose that there are two time steps ($t_{\text{max}} = 2$). Searching in the second time step will yield an expected reward of $\mu$. Settling for anything less than $\mu$ after the first time step would therefore result in a lesser expected reward in the second time step. 

We also prove that the optimal threshold ($T$) increases in the search time ($t_{\text{max}}$), and is finite as long as the search time is finite (Appendix~\ref{app:derivations}). Again, both results are intuitive: more search time increases the likelihood and cumulative payoff of achieving an ambitious target.

\subsection{Overshooting the optimal threshold is costlier than undershooting} An agent overshooting the optimal threshold is expected to receive lower rewards than an agent undershooting the optimal threshold by the same amount (Figs. \ref{fig:theory-vs-sim}a, \ref{fig:smooth_skew}, \ref{fig:group_upward}, and \ref{fig:search_costs}). This implies that uncertainty over the reward distribution should reduce the optimal satisfaction threshold (i.e., it should motivate caution), analogous to the way that left-skewed uncertainty motivates risk aversion in other contexts \cite{benuzzi2024skewness}. However, in settings where agents learn about the reward distribution through searching and they can adjust their satisfaction thresholds accordingly, Kohn and Shavell \cite{kohn1974theory} show that uncertainty can increase the optimal satisfaction threshold due to the information value of searching. 

Fig.~\ref{fig:theory-vs-sim}b and c illustrate two real-world contexts in which people seem to behave as though they understand the search tradeoff that our theory captures. Fig.~\ref{fig:theory-vs-sim}b compares the frequencies with which heterosexual men and women in four U.S. cities sent messages to prospective partners on a dating app, as a function of their prospective partners' desirability compared to their own desirability (data from ref. \cite{bruch2018aspirational}). Long-term partners tend to be similar in their desirability \cite{george2015couple}. Thus, a zero desirability gap might be analogous to the mean reward ($\mu$) in our model. In the online dating market, both men and women expend the greatest (modal) search effort on prospective partners slightly more desirable than themselves, as our theory would predict (Fig.~\ref{fig:theory-vs-sim}b). 

Fig.~\ref{fig:theory-vs-sim}c shows the distribution of college applications from the 2008 U.S. high-school graduating class, comparing students' own scores on the Scholastic Aptitude Test (SAT) to those of the median current student at the college they were applying to (data from ref. \cite{hoxby2012missing}). Attending a college where one has the median SAT score might be analogous to the mean reward ($\mu$) in our model. As in the dating example, the modal search (application) effort occurred near this mean college. 

However, unlike the dating example, students sent more applications to weaker colleges than they sent to stronger ones. There are at least two reasons for this. First, Hoxby \cite{hoxby2012missing} showed that most applications in this left tail came from low- and middle-income students, who faced greater non-academic constraints (e.g., geographic, economic) on which colleges they could attend. Second, unlike agents in our model, college applicants pursue multiple strategies (applications) simultaneously rather than sequentially (unless they transfer). They may thus prefer to include less-ambitious schools in their application portfolio to avoid striking out or having to search again the following year---the `safety school' strategy. Despite these nuances, the basic qualitative tradeoff our model illustrates remains: applicants who focus on or choose safety schools risk losing out on higher-value options; applicants who focus on out-of-reach schools risk wasting their time and money or striking out. As a result, applicants focus most attention on schools they are well matched to (Fig. ~\ref{fig:theory-vs-sim}c). 

\subsection{Reward landscape left skewness and ruggedness increase optimal satisfaction thresholds}

For a landscape with smoothness $\varphi$, the variance of the associated AR(1) process is $\text{var}[X_t] = (1+\varphi)/(1-\varphi)$. The analytical expression for the expected cumulative reward on smooth landscapes can therefore be derived by simply scaling the threshold $T$ by a factor of $1/\sqrt{\text{var}[X_t]}$ in equation \eqref{eq:expected-reward}. 

\begin{figure}[t!]
\centering
\includegraphics[width=\linewidth]{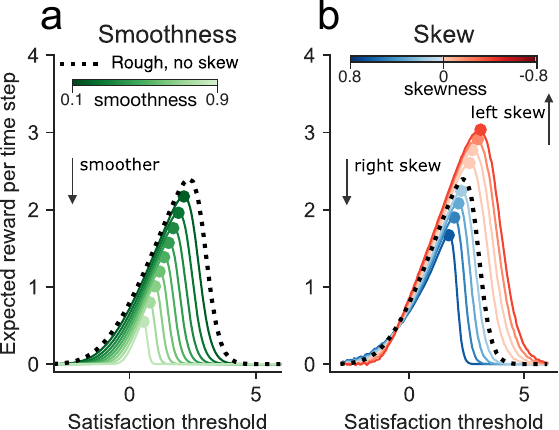}
\caption{\textbf{Effects of smoothness and skewness.} Smoother reward landscapes yield lower cumulative rewards than the maximally rugged, non-skewed landscapes (a). Left-skewed reward landscapes yield higher cumulative rewards (b). Smooth landscapes are generated analytically. For the skewed landscape simulations, $\epsilon_t \sim \mathcal{SN}(0, 1,\alpha)$, $\varphi=0$, $t_{\text{max}}=1000$, where the parameter $\alpha$ is varied to result in skew values from $-0.8$ to $0.8$, each averaged over $10^4$ simulations.}
\label{fig:smooth_skew}
\end{figure}

Fig.~\ref{fig:smooth_skew} shows how varying smoothness (or ruggedness) and skewness affects the optimal satisfaction threshold, and the expected reward per time step as a function of the satisfaction threshold. 

Rugged landscapes create higher-variance rewards, which increases achievable expected rewards and optimal satisfaction thresholds (Fig.~\ref{fig:smooth_skew}a). Thus, agents should be more ambitious on rugged landscapes. The reward distribution for the first time period is the same for smooth and rugged landscapes, but autocorrelation reduces the subsequent variance on smooth landscapes. This means that agents can only make incremental changes: setting their sights too high will likely result in never encountering the desired reward, which would require a gradual climb due to the aurocorrelation. 

Holding the mean reward ($\mu$) constant, left (right) skew implies a higher (lower) median and mode reward (Fig.~\ref{fig:conceptual_figure}). Left-skewed reward landscapes therefore have higher optimal satisfaction thresholds, compared to the mean ($\mu$), since higher rewards are more abundant (Fig.~\ref{fig:smooth_skew}b). Expected rewards are also higher in left-skewed reward landscapes, at the optimal threshold. Right-skewed reward landscapes have the opposite properties. Simulations in Fig.~\ref{fig:smooth_skew}b assume skew-normal reward distributions, with mean of zero and variance of one, as in other simulations. 

\subsection{Social comparison hurts performance and penalizes ambition}

We have previously assumed that agents know the reward distribution. What if they instead estimate the distribution, and evaluate their satisfaction, by looking at peers' rewards \cite{goethals1977social}? This type of behavior is called social comparison \cite{suls2002social,schulz2015reference,levy2004reported,bygren2004pay}. Indeed, in reality people often focus specifically on peers receiving higher rewards than they are---``upward social comparison'' \cite{gerber2018social, muller2010being}. 

We model cohort comparison and upward social comparison. Cohort comparison assumes that each agent knows the mean and variance of their peers' rewards (across all peers) at the previous time step. Upward social comparison assumes that each agent makes these estimates only taking into account their own reward and peers' rewards that were greater. Note that when agents define their satisfaction threshold relative to the performance of others, a previously satisfied agent can become unsatisfied and return to searching if group performance changes compared to their threshold.

\begin{figure}
\centering
\includegraphics[width=\linewidth]{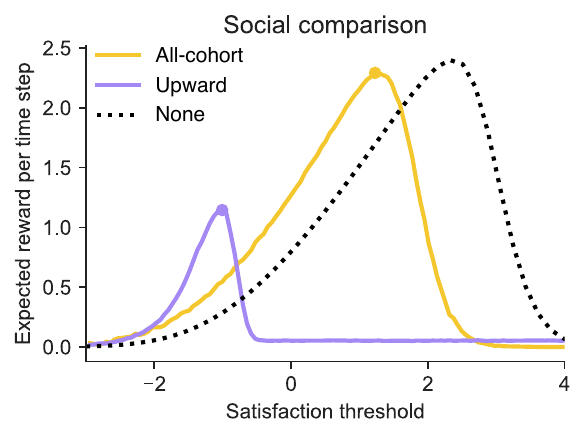}
\caption{\textbf{Social comparison is costly.} Mean rewards for cohort comparisons and upward social comparisons are shown on a rough, non-skewed landscape. Cohort comparison (yellow curve) lowers the optimal satisfaction threshold and cumulative reward. Upward social comparison (purple curve) further lowers the optimal threshold and substantially hinders performance. Both the expected reward and the optimal satisfaction threshold are lower compared to the reference landscape (black, dashed curve), which assumes no social comparison. Parameter values are: $\epsilon_t \sim \mathcal{N}(0,1)$, $\varphi=0$, and $t_{\text{max}}=1000$. Cohorts include $100$ agents, with thresholds uniformly sampled from $[-3,6]$. Results shown are averaged over $10^4$ simulations.}
\label{fig:group_upward}
\end{figure}

Fig.~\ref{fig:group_upward} shows simulations for 100-agent cohorts. Agents vary in their satisfaction thresholds, each chosen randomly from a uniform distribution. The optimal satisfaction threshold is slightly lower with cohort comparison than when agents know the true reward distribution. Because agents can become unsatisfied if the cohort statistics change, setting a lower threshold prevents agents from overreacting to others' success. Additional cohort sizes and threshold intervals are shown in Fig.~\ref{fig:cohort-size} and Fig.~\ref{fig:threshold-limits} in Appendix~\ref{appendix:robustness}.

Upward social comparison substantially lowers agents' expected rewards, and makes the optimal threshold lower than the perceived mean (Fig.~\ref{fig:group_upward}). The perceived mean reward is higher than the true mean ($\mu$), due to the upward social comparison. Cohorts of agents engaging in upward social comparison end up never satisfied when they set satisfaction thresholds above the mean of those doing better than them. In other words, upward social comparison sets agents up for disappointment and failure by causing them to ignore useful information about the true reward distribution (from lower-performing agents) and by creating unrealistic expectations.

\subsection{Search costs}
\label{sec:search-costs}
To consider the effects of search costs, we assume that the agent pays a constant cost, $c$, in each period they spend searching for a new strategy. Inserting this assumption into equation \eqref{eq:1}, the agent's reward in time period $t$, $X_t$, becomes:

\begin{equation}
X_t=
\begin{cases}
\varphi X_{t-1}+(1-\varphi)\epsilon_t-c & X_{t-1}<T,\\ 
X_{t-1}  & X_{t-1} \geq T.
\end{cases}
\label{eq:3}
\end{equation}

From equation \eqref{eq:expected-reward}, their expected reward becomes:

\begin{align}
\mathbb{E}[\text{reward}] &= 
\nonumber \\ \sum_{t_x=0}^{t_{\text{max}}} \Bigg[ (\mu_{\text{explore}}(T)-c) t_x + 
\mu_{\text{exploit}}(T) \left( t_{\text{max}} - t_x \right) \Bigg] 
\nonumber \\ \times (1 - \Phi(T)) \Phi(T)^{t_x}.
\label{eq:expected-reward-cost}
\end{align}

\begin{figure}
\centering
\includegraphics[width=\linewidth]{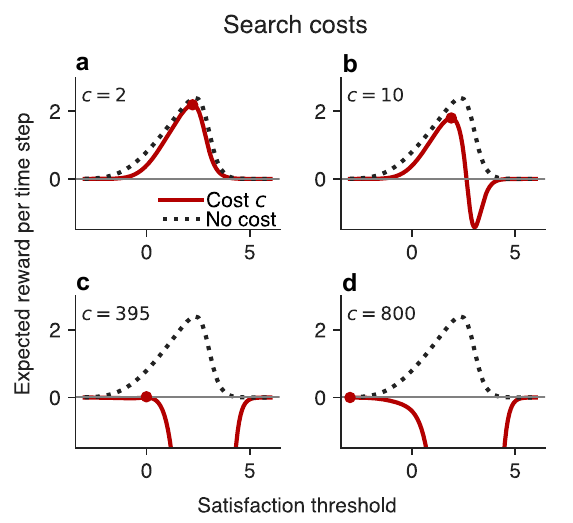}
\caption{\textbf{Effects of search costs.} Expected rewards for search with a per-timestep search cost on a rough, non-skewed landscape. The cost values are: (a) $c=2$, (b) $c=10$, (c) $c=395$, (d) $c=800$. Introducing a search cost (colored curves) lowers the optimal satisfaction threshold and can produce negative expected rewards when either the cost or the threshold are too high. 
In all cases, both the expected reward and the optimal satisfaction threshold are lower than those of the reference landscape (black dashed curve). The sums are computed numerically based on equation~\eqref{eq:expected-reward-cost}. Parameter values are: $\epsilon_t \sim \mathcal{N}(0,1)$, $\varphi=0$, and $t_{\text{max}}=1000$.}
\label{fig:search_costs}
\end{figure}

Fig.~\ref{fig:search_costs} shows the effects of increasing the search costs ($c$) on the expected reward as a function of the satisfaction threshold ($T$). Search costs gradually reduce both the optimal satisfaction threshold and the maximum achievable reward (Fig.~\ref{fig:search_costs}a). Search costs increase the degree to which being too ambitious is costlier than being too cautious, such that being too ambitious can now make expected rewards worse than the expected rewards of never searching (i.e., $\mu$ per time step) (Fig.~\ref{fig:search_costs}b). As long as a satisfaction threshold exists with higher expected rewards than never searching, however small, the optimal satisfaction threshold will be still be higher than the mean of available rewards ($\mu$) (Fig.~\ref{fig:search_costs}c). Otherwise, the optimal threshold will be $-\infty$ (i.e., the agent should never search) (Fig.~\ref{fig:search_costs}d).

In other words, search costs only affect our main qualitative result---that optimal ambition targets finite rewards strictly larger than the mean of available rewards---in the extreme case where costs are so high that they prohibit all search.

\subsection{Applications}
\begin{figure*}[t!]
\centering
\includegraphics[width=0.9\linewidth]{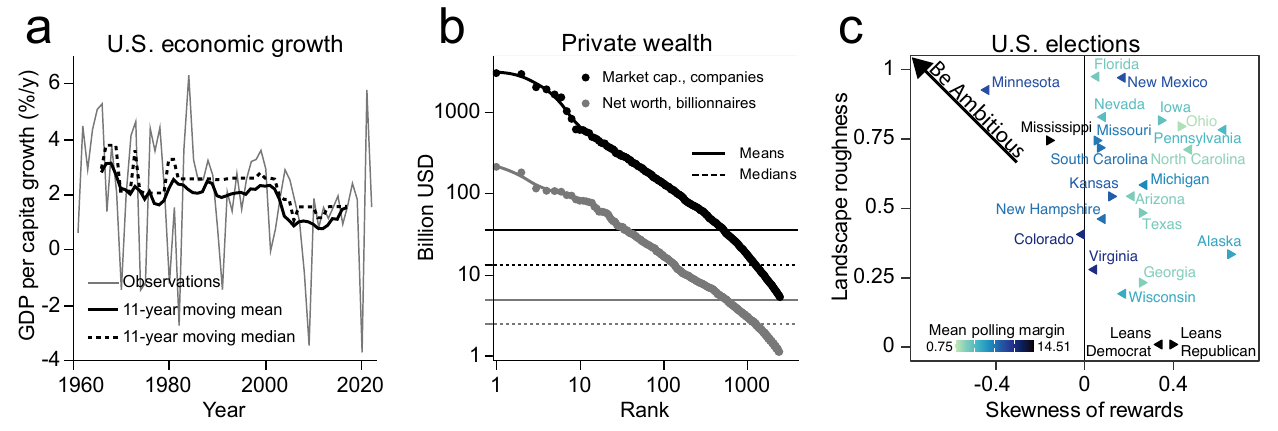}
\caption{\textbf{Possible applications}. Panel a: U.S. growth rates in gross domestic product (GDP) per capita (gray) \cite{owidgrowth}, compared moving means (black) and medians (dashed). The medians are greater than the means, indicating left skew. Panel b: distributions of private wealth among the 2500 top companies \cite{kagglecos} and billionaires \cite{kagglebillionaires}. The means are much greater than the medians, indicating right skew. Panel c: monthly polling margins, from the perspective of Republicans (i.e. Republican share $-$ Democrat share) in U.S. swing states during 2020, prior to the presidential election \cite{fivethirtyeight}. Darker colors indicate larger average differences between the two parties. Lighter-colored states have smaller average margins.}
\label{fig:empirical}
\end{figure*}

Although our model is stylized and does not literally describe real-world decision-making in its full complexity, Fig.~\ref{fig:empirical} illustrates examples of where our model's qualitative insights regarding smoothness and skewness could be applied to economic policy, entrepreneurship and political campaigns. These examples reveal a subtle but important distinction between optimal ambition and optimal risk-taking, which should generalize beyond our stylized setting. It follows directly from the statistical properties of the reward distribution, and not from the specific search procedure assumed.

Economic growth rates tend to have left-skewed distributions in developed countries, because recessions tend to have larger magnitudes than booms \cite{burgess2020optimistically}. Fig.~\ref{fig:empirical}a shows this pattern in U.S. data. From a policy standpoint, the left-skewed distribution of economic growth rates implies that risk taking is costly. If downside risks are larger than upside risks, risk-taking lowers the expected outcome. However, our model implies seemingly opposite advice for \textit{ambition}, compared to the mean reward. A country's mean economic growth rate is dragged down by large and rare recessions (Fig.~\ref{fig:empirical}a). Policymakers would therefore be unwise to be satisfied with growth rates close to this mean in a typical year. In other words, the left-skewness of economic growth rates calls for less risky but more ambitious (compared to the mean) economic policymaking.

Private wealth distributions have the opposite property: they are heavily right-skewed \cite{scheffer2017inequality}. Fig.~\ref{fig:empirical}b illustrates the wealth distributions of the world's richest billionaires and the world's largest companies. A right-skewed distribution of private wealth outcomes implies that upside risks are often larger than downside risks---risk-taking in entrepreneurship often means either winning big or losing small. Consequently, risk taking is often beneficial in business and private investment \cite{benuzzi2024skewness}. However, our model suggests that the right skewness of private wealth calls for \textit{less} ambition, compared to the mean available reward. This is counterintuitive, but it is related to the fact that mean success in wealth building and entrepreneurship is highly influenced by rare and extremely successful `unicorns'. Thus, an entrepreneur or investor who sets a high satisfaction threshold compared to the mean may miss opportunities for important and achievable successes. 

Fig.~\ref{fig:empirical}c shows the skewness and smoothness of monthly polling margins in swing states ahead of the 2020 U.S. presidential election \cite{fivethirtyeight}, from the perspective of the Republican party. (For the Democrats, skewness would have the opposite sign and smoothness would be the same.) Parties and candidates undertake platforms, messaging, and on-the-ground campaign strategies that either explicitly or implicitly aim for certain vote margin targets in different states and regions. Higher margins are more likely in states with more rugged and left-skewed polling distributions. Therefore, our model implies that parties should pursue more ambitious (albeit less risky, given the left-skew) political strategies in such states, all else being equal.   

For example, pre-election polling distributions in 2020 would have suggested that Republicans had a higher chance of outperforming their mean polling margins in Minnesota and South Carolina than they did in neighboring Wisconsin and Georgia (Fig.~\ref{fig:empirical}c). This suggests that the Republicans had a greater incentive to pursue ambitious campaign strategies in these states, all else equal. 

Of course, there are many other factors, besides expected achievable vote margins, that are relevant in allocating campaign resources to states in practice. Such factors include (but are not limited to) the Electoral College rewards, tradeoffs and synergies between strategies targeting different regions and constituencies, and how close the expected vote margins are to zero, given the first-past-the-post system in most states. We also note that pre-election polling distributions are used here as a proxy for the distribution of available rewards (especially its smoothness and skewness), not as a literal stand-in for the stream of rewards over time. It is the election-day vote that matters in practice, of course.

\section{Discussion}

Our model formalizes the folk intuition that optimal ambition is both finite and aims to do better than average (Fig.~\ref{fig:theory-vs-sim}a). Optimal ambition must balance the risks of being under-ambitious and over-ambitious, which can both lead to missing out on higher attainable rewards. We show empirical examples (from dating and college applications) in which people seem to behave as though they understand this balance, either consciously or sub-consciously (Fig.~\ref{fig:theory-vs-sim}b,c). Although our model assumes that all agents face the same reward distribution, the available rewards---and consequently, the optimal ambition---varies widely across people and contexts in reality. 

Our model then adds more precise insights. First, we show that being too unsatisfiable is costlier than being too easily satisfied, on average (Fig.~\ref{fig:theory-vs-sim}a). Second, we show that left-skewed and rugged reward distributions should motivate more ambition (Fig.~\ref{fig:smooth_skew}). This exposes a subtle difference between optimal ambition and optimal risk taking, as left-skewed rewards motivate \textit{less} risk-taking \cite{benuzzi2024skewness}. Third, we show that upward social comparison is costly by creating unrealistic expectations (Fig.~\ref{fig:group_upward}). Fourth, we show that our model's main insight is insensitive to search costs, up the point where search costs disincentivize all search (Fig. ~\ref{fig:search_costs}). We describe how these insights apply qualitatively to real-world contexts relating to economic policy, wealth building and political campaigns (Fig.~\ref{fig:empirical}). We discuss additional real-world examples in the introduction.  

Our model is admittedly simpler than real-world decision contexts, including those we explore empirically. Indeed, our model's accessibility to a broad audience is one of its contributions. However, `explore-exploit' and sequential searches are common abstractions in areas such as microeconomics \cite{kohn1974theory}, management science \cite{lavie2010exploration,greve2007exploration}, harvesting \cite{richard2018commercial} and animal foraging \cite{macarthur1966optimal, cowie1977optimal}. Our `satisfaction' framework is somewhat related to the "satisficing" concept in behavioral economics \cite{simon1955behavioral}, but it is also distinct in that our threshold is optimized, rather than being a behavioral heuristic. Yang et al. \cite{yang2020us} provide a recent example of modeling satisficing in the context of   political polarization.

The framework of optimal search has been previously studied through different lenses in different disciplines, some of which provide insights into how our results might extend to more complex settings. For example, Kohn and Shavell \cite{kohn1974theory} consider how time preference, risk-averse utility and uncertainty affect optimal search. Their model is too complex for optimal satisfaction thresholds to be derived in general, but they show that risk-averse utility and discounting future rewards would each reduce optimal satisfaction thresholds. These insights are intuitive. They also show that uncertainty can raise optimal satisfaction thresholds by increasing the information value of searching \cite{kohn1974theory}. This finding is somewhat counterintuitive in light of our finding that higher-than-optimal thresholds are costlier than lower-than-optimal thresholds (Fig.~\ref{fig:theory-vs-sim}a), combined with that fact that left-skewed uncertainty typically motivates risk avoidance \cite{benuzzi2024skewness}. 

From a complementary perspective, Yin et al.~\cite{yin2019quantifying} examine the dynamics of learning from failure in a model where agents choose between incremental improvement and exploring drastically different opportunities, determined by a single parameter controlling how many previous attempts are incorporated into learning. They find a phase transition between regimes of incremental refinement and disjointed search. Within our framework, the ability to refine and improve the chosen strategy over time would likely lower the optimal satisfaction threshold.

Applied mathematicians have widely studied `the secretary problem', where one seeks to maximize the chance of selecting the best option from a sequentially observed set \cite{freeman1983secretary}. For example, MacQueen and Miller \cite{macqueen1960optimal} characterize the optimal search duration under these conditions. The exact bounds and optimal strategies depend on the underlying distribution and corresponding assumptions. In the case of AR(1) processes (like our model), several bounds have been derived for exponentially distributed steps ~\cite{novikov2009distributions, christensen2012phase, christensen2011elementary}. However, because of their complexities and sensitivities to assumptions, empirical applications often rely on powerful but opaque machine learning approaches~\cite{becker2019deep, ciocan2022interpretable}.

Social comparisons have been theoretically predicted and empirically documented to affect decision-making in numerous human and animal contexts. Humans evaluate their successes and status relative to their peers  \cite{layard2006happiness,layard2011happiness}, and, as a result, they make consumption decisions aimed at optimizing both absolute and relative outcomes (e.g., ref. \cite{arrow2009conspicuous}). Human fact-finding and decision-making processes can be either enhanced (e.g., the wisdom of crowds \cite{surowiecki2005wisdom}) or corrupted (e.g., groupthink \cite{janis1971groupthink}) by social forces. In animals, instincts for social conformism have also been linked to improved (e.g., ref. \cite{berdahl2013emergent}) and diminished (e.g., ref. \cite{dussutour2007group}) decision-making in different contexts. 

Future research should explore other complexities that our model does not consider. For example, there can be inherent benefits in reality to sticking with one strategy (e.g., accruing expertise, deepening a relationship, refining an innovation \cite{yin2019quantifying}). Similarly to search costs (e.g., disruptions from moving cities or changing jobs), this would likely reduce optimal satisfaction thresholds. Real people are also prone to weighing potential losses more than potential gains (known as `loss aversion' \cite{kahneman1991endowment}). This, too, would reduce the optimal satisfaction threshold in absolute terms in our model, by reducing the utility that agents assign to low-end nominal rewards.  

On the other hand, in cases such as foraging or natural resource extraction, rewards can be depleted over time. This could result in decreasing expected cumulative rewards and lower optimal satisfaction thresholds, and/or cycles of exploration and exploitation. However, Burgess et al.~\cite{burgess2020opportunities} show that fishers still maximize their catch by revisiting patches with slightly above-average expected rewards, similar to our model's prediction. Depletion merely changes which patches these are over time. In social settings, reward depletion can create complex interactions, such as in the example of `dynamic foraging'~\cite{bidari2022stochastic}. In addition, social dynamics can encourage `optimal distinctiveness', where optimal ambition is effectively capped by the need to maintain group identity~\cite{Brewer2003-BREODS}. 

Our agents are simplified in terms of both their decisions and their inferences about available rewards. Our social comparison model does not consider structured social networks. We model random rather than intelligent search processes, though it is straightforward to conjecture that intelligent search would increase optimal ambition by making high but rare rewards easier to find. There are many other worthwhile extensions to our framework. Finally, we operationalize ambition as the reward satisfaction threshold. Alternative frameworks could more explicitly link ambition to a combination of target reward and other factors such as willingness to pay search costs, or willingness to search on a rugged, high-variance landscape.

Despite these limitations, our conceptual results are intuitive and likely to generalize. For example, increasing ambition makes satisfactory rewards harder to find under any search process. Making high rewards more abundant---via skewness or ruggedness---makes ambitious targets easier to achieve.  Upward social comparison distorts perceptions of reality, which hinders rational decision making. Search costs reduce optimal ambition, but optimal ambition still generally exceeds the mean of available rewards.

How ambitious to be is an important question in a wide range of professional, political and personal contexts. Our model, results and examples offer entry points for mathematically precise, but also accessible and intuitive, explorations of this question.

\section*{Data availability}
For parts of this work, we used previously published data, available from refs.  \cite{bruch2018aspirational,hoxby2012missing,fivethirtyeight,owidgrowth,kagglecos,kagglebillionaires}.

\section*{Code availability}
Our simulation code is available on GitHub at 
 \href{https://github.com/kathlandgren/ambition}{https://github.com/kathlandgren/ambition}. The simulation algorithm is described in Appendix~\ref{app:algorithm}.

\section*{Acknowledgements}

We thank members of the Burgess lab, Ellen DeGennaro, Todd Cherry, Jonas L. Juul, Chase Thiel, and Scott Beaulier for comments. E.L. acknowledges funding from the Cooperative Institute for Research in Environmental Sciences (CIRES) Visiting Fellows Program at the University of Colorado Boulder, funded by NOAA Cooperative Agreement NA22OAR4320151.

\appendix
\section{Additional derivations}
\label{app:derivations}

\subsection{Limiting behavior as $t_{\text{max}} \rightarrow \infty$}

Recall that---in the special case of a rugged landscape with Gaussian rewards---we can write down the expected cumulative reward for a given agent by summing over all the possible lengths of the exploration phase (equation \eqref{eq:expected-reward}).

By substituting the expressions for the inverse Mills ratios $\mu_{\text{explore}}$ and $\mu_{\text{exploit}}$, we can simplify this expression as follows:  

\begin{align}
\mathbb{E}[\text{reward}] &= t_{\text{max}} \phi(T) \sum_{t_x=0}^{t_{\text{max}}} \Bigg[ \Phi^t(T) - \Phi^{t_x-1}(T) \dfrac{t_x}{t_{\text{max}}} \Bigg].
\label{eq:simplified-reward}
\end{align}

It is convenient to consider the expected reward per time step, $\mathbb{E}[\text{reward}]/t_{\text{max}}$. We can now separate the expression into two separate sums: 
\begin{equation}
\dfrac{\mathbb{E}[\text{reward}]}{t_{\text{max}}}= \phi(T) (S_1+S_2),
\label{eq:reward-per-tstep}
\end{equation}
which we consider separately. Here,  $S_1 =  \sum_{t_x=0}^{t_{\text{max}}}  \Phi^t_x(T) $ is a geometric sum, and $S_2 =\sum_{t_x=0}^{t_{\text{max}}} \Phi^{t_x-1}(T) \dfrac{t_x}{t_{\text{max}}}$ is an arithmetico-geometric sum. We can write down the partial sum expressions for both $S_1$:

\begin{align}
S_1 &=  \sum_{t_x=0}^{t_{\text{max}}}  \Phi^t(T) = \dfrac{1-\Phi^{t_{\text{max}}+1}(T)}{1-\Phi(T)},
\label{eq:S1}
\end{align}

 and $S_2$:

\begin{align}
S_2 &=  \sum_{t_x=0}^{t_{\text{max}}}  \Phi^{t_x-1}(T) \dfrac{t_x}{t_{\text{max}}}
\\ &= \dfrac{1}{t_{\text{max}}} \Bigg[ \dfrac{1-(t_{\text{max}}+1)\Phi^{t_{\text{max}}}(T)}{1-\Phi(T)} + \dfrac{\Phi(T)(1-\Phi^{t_{\text{max}}}(T)}{(1-\Phi(T))^2} \Bigg].
\label{eq:S2}
\end{align}

Now we can consider infinite sums, which allow us to describe the behavior of the expected reward in the limiting case as $t_{\text{max}} \rightarrow \infty$: 

\begin{align}
\lim_{t_{\text{max}} \rightarrow \infty} S_{1} &=  \dfrac{1}{1-\Phi(T)}
\label{eq:S1infty}
\end{align}

and 

\begin{align}
\lim_{t_{\text{max}} \rightarrow \infty}S_2 &=  \dfrac{1}{t_{\text{max}}}\left[ \dfrac{1}{1-\Phi(T)}+\dfrac{\Phi(T)}{1-\Phi^2(T)}\right]=0.
\label{eq:S2infty}
\end{align}

Now, substituting  \eqref{eq:S1infty} and \eqref{eq:S2infty} into \eqref{eq:reward-per-tstep}, we obtain the following:

\begin{align}
& \lim_{t_{\text{max}} \rightarrow \infty}\frac{\mathbb{E}[\text{reward}]}{t_{\text{max}}} \nonumber \\
&= \lim_{t_{\text{max}} \rightarrow \infty} \phi(T) \ \times \nonumber\\
& \left[ \frac{1}{1-\Phi(T)} 
    - \frac{1}{t_{\text{max}}} \left( \frac{1}{1-\Phi(T)}
    + \frac{\Phi(T)}{1-\Phi^2(T)} \right) \right] \nonumber \\
&= \dfrac{\phi(T)}{1-\Phi(T)} \nonumber \\
&= \mu_{\text{exploit}}(T).
\label{eq:more-simplified-reward}
\end{align}

This result implies that given infinite time, the ``explore'' period is negligible compared to the ``exploit'' stage, and the agent can be arbitrarily ambitious.

\subsection{The optimal satisfaction threshold is finite and increases as the total time increases}
For any threshold, $T$, the probability of finding a strategy that satisfies the threshold increases with the length of the search, $t_{\text{max}}$. The amount of time one gets to exploit a satisfactory strategy also increases in $t_{\text{max}}$, all else equal. Both of these patterns shift incentive towards a larger optimal threshold as $t_{\text{max}}$ increases.  In the limiting case of $t_{\text{max}}\rightarrow \infty$, the expected reward per time step is equal to $\mu_{\text{exploit}}$, 
meaning that when time is infinite, agents can be arbitrarily ambitious, but optimal ambition is finite in finite time.

\subsection{The optimal satisfaction threshold is strictly greater than the mean reward}
\label{sec:finite-optimum}

The reward in the `exploit' phase ($\mu_{\text{exploit}}(T)$) is always positive and increases in $T$. The expected number of time steps it takes to satisfy threshold $T$ is $\frac{1}{1-\Phi(T)}$: at $T=0$, the expected length of the exploration phase is two time steps, and it increases exponentially as the threshold $T$ increases. Intuitively then, agents who can afford the exploration phase to be longer than two time steps should target positive thresholds $T$ (i.e., thresholds larger than the mean, $\mu = 0$).

\begin{figure}[tb]
\centering
\includegraphics[width=\linewidth]{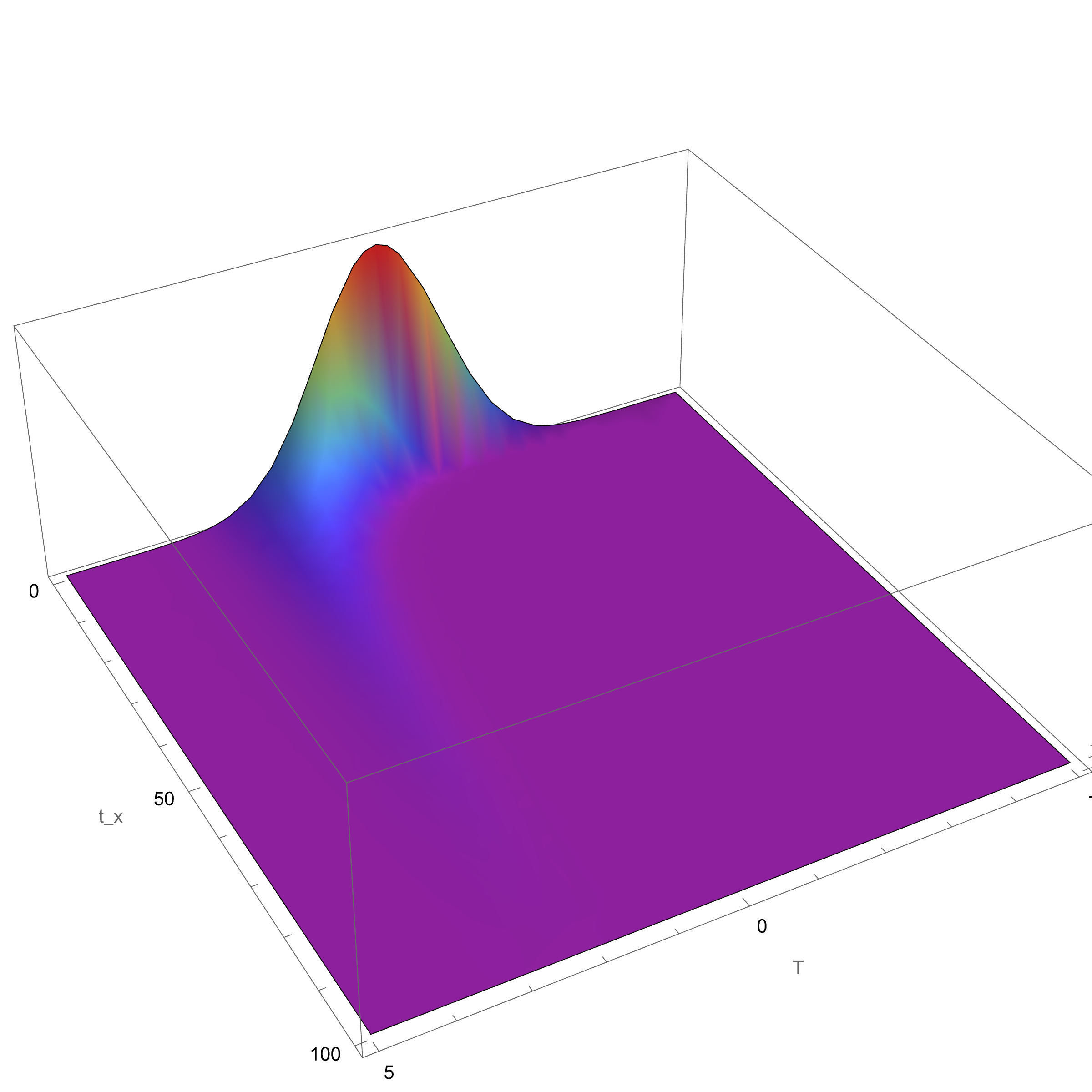}
\caption{The summands in equation \eqref{eq:simplified-reward} as a function of threshold $T$ and time step $t_x$. Each summand can be thought of as a slice for a fixed time $t_x$. Note that the maximum of each summand occurs at $T>0$ (i.e., at a satisfaction threshold above the mean, $\mu = 0$).}
\label{fig:summand}
\end{figure}

The reward in the ``exploit'' phase $\mu_{\text{exploit}}(T)$ is always positive and increases in $T$, but is close to zero for negative values of threshold $T$, so high positive thresholds yield higher rewards and negative thresholds tend to yield rewards close to the mean. 
The expected number of time steps it takes to satisfy threshold $T$ is $\dfrac{1}{1-\Phi(T)}$. At $T$=0, the expected length of the exploration phase is two time steps, and grows exponentially as threshold $T$ increases. 
Intuitively, the agents who can afford the exploration phase to be longer than two time steps should target positive thresholds $T$.

We show that the optimal threshold is above the mean for the maximally rugged landscape. We can split the summand in equation \eqref{eq:expected-reward} into the reward component \begin{equation}
R(T,t_x)=\mu_{\text{explore}}(T) t_x + \mu_{\text{exploit}}(T) \left( t_{\text{max}} - t_x \right),
\label{eq:reward-comp}
\end{equation}

and the probability component 

\begin{equation}
P(T,t_x)=(1 - \Phi(T)) \Phi(T)^{t_x}.
\label{eq:prob-comp}
\end{equation}

\subsubsection{The reward component increases near-linearly in $T$}

As the threshold $T\rightarrow \infty$, $\mu_{\text{explore}}\rightarrow 0$, and $\mu_{\text{exploit}}\rightarrow T$.
As the threshold $T\rightarrow -\infty$, $\mu_{\text{explore}}\rightarrow T$, and $\mu_{\text{exploit}}\rightarrow 0$ \cite{small2010expansions}. For a given $t_x$, as $T$ gets large, the linear combination of $\mu_{\text{explore}}$ and $\mu_{\text{exploit}}$ grows near-linearly in $T$.

\subsubsection{The probability component \eqref{eq:prob-comp} is unimodal}

The partial derivative of the probability component $P(T,t_x)$ with respect to threshold $T$ is 

\begin{equation}
\dfrac{\partial P}{\partial T} = t_x f(T) \left[ (1-\Phi(T))\Phi(T)^{t_x-1}-\Phi(T) \right].
\end{equation}
The probability component $P(T,t_x)$ has a single critical point at $\Phi(T)=\frac{t_x}{t_x+1}$.
Since the second derivative $\frac{\partial^2 P}{\partial T^2}<0$, the corresponding value of $T$ maximizes $P(T,t_x)$ for a fixed $t_x$.

\subsubsection{The case of $t_x=0$}
For the case $t_x=0$, the summand (the product of the reward component $R(T,t_x)$ and the probability component $P(T,t_x)$ reduces to $t_{\text{max}}\phi(T)$, which achieves its maximum at $T=0$.

\subsubsection{Optimal threshold is greater than the mean}
Since the reward component $R(T,t_x)$ increases near-linearly in $T$, and $P(T,t_x)$ is maximized when $\Phi(T)=\frac{t_x}{t_x+1}$, the summand $R(T,t_x)P(T,t_x)$ for a given $t_x$ achieves its maximum for a threshold $T^*_{t_x}>\Phi^{-1}\left( \frac{t_x}{t_x+1} \right)$ (see Fig.~\ref{fig:summand}).

As $t_x$ increases, $\frac{t_x}{t_x+1}$ increases also, and therefore $T^*_{t_x}$ increases, meaning that each subsequent summand achieves its maximum at a greater threshold value $T$ than the one for a previous summand. 

\begin{figure*}[tbhp]
\centering
\includegraphics[width=1\linewidth]{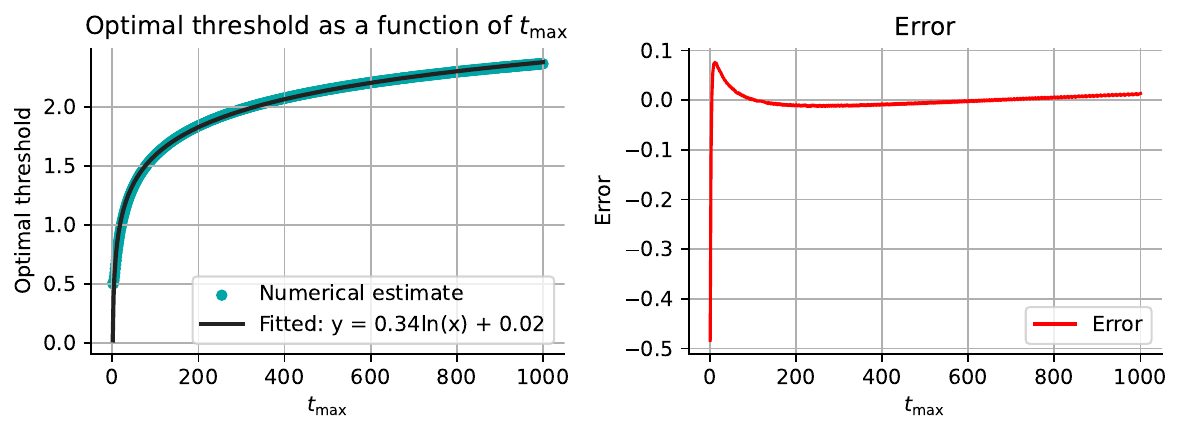}
\caption{The optimal threshold, $T^*$, is closely approximated by a logarithmic function. The left panel shows the numerical estimate and the fitted logarithmic curve. The right panel shows the error $T^*-y$. Note that the error changes sign, meaning that the optimal threshold grows faster than the logarithmic fit.}
\label{fig:opt_thresh}
\end{figure*}

Let $t_x=1$. The maximizing satisfaction threshold $T^*_{1}=\Phi^{-1}\left( \frac{1}{2} \right)=0.$ Since every summand $R(T,t_x)P(T,t_x)$ is unimodal, every summand for $t_x>1$ achieves its maximum for $T^*_{t_x}>0$, which implies that $\left.\frac{\partial RP(T,t_x)}{\partial T} \right\vert_{T=0}>0$, $\forall t_x>1$. 
Therefore for $t_{\text{max}}>1$, the optimal threshold is greater than the mean. Fig.~\ref{fig:opt_thresh} shows that while the optimal threshold can be approximated by a logarithmic function, it grows faster than the logarithmic fit.

\clearpage
\section{Simulation Algorithm}
\label{app:algorithm}

\begin{figure}[!ht]
\centering
\begin{minipage}{\linewidth}
\begin{algorithmic}[1]
\small
\State \textbf{Input:} $t_{\max}$, $N$ agents, threshold $T$, information type \texttt{infotype}, landscape mean $\mu$, variance $\sigma^2$, smoothness $\varphi$, skew $\alpha$
\State \textbf{Output:} matrix \texttt{reward} of size $t_{\max}\times N$
\Statex

\Comment Initialize the matrix of rewards
\State \texttt{reward} $\gets$ \texttt{matrix}(nrow $= t_{\max}$, ncol $= N$)

\Comment Assign random initial rewards
\For{$a \gets 1$ \textbf{to} $N$}
  \State draw $\epsilon_0 \sim \mathcal{SN}(\mu,\sigma,\alpha)$
  \State \texttt{reward}$[0,a] \gets \epsilon_0$
\EndFor
\Statex

\Comment Loop through time
\For{$t \gets 1$ \textbf{to} $t_{\max}$}

  \Comment Loop across agents
  \For{$a \gets 1$ \textbf{to} $N$}

    \Comment Summary stats per \texttt{infotype}
    \If{\texttt{infotype} $=$ \texttt{"cohort"}}
      \State \texttt{info} $\gets \{\,\texttt{reward}[t-1,j] : j\neq a\,\}$
      \State $\mu_{\text{info}} \gets \texttt{mean}(\texttt{info})$, \quad $\sigma_{\text{info}} \gets \texttt{sd}(\texttt{info})$
    \ElsIf{\texttt{infotype} $=$ \texttt{"upward"}}
      \State \texttt{info} $\gets \{\,\texttt{reward}[t-1,j] : \texttt{reward}[t-1,j] > \texttt{reward}[t-1,a]\,\}$
      \State $\mu_{\text{info}} \gets \texttt{mean}(\texttt{info})$, \quad $\sigma_{\text{info}} \gets \texttt{sd}(\texttt{info})$
    \Else  \Comment \texttt{"landscape"}
      \State $\mu_{\text{info}} \gets \mu$, \quad $\sigma_{\text{info}} \gets \sigma$
    \EndIf

    \Comment Compare to threshold and move or stay
    \State \texttt{prev} $\gets$ \texttt{reward}$[t-1,a]$
    \If{\texttt{prev} $< \mu + T\,\sigma$}
      \State draw $\epsilon_t \sim \mathcal{SN}(\mu,\sigma,\alpha)$
      \State \texttt{new} $\gets \varphi \cdot \texttt{prev} + (1-\varphi)\cdot \epsilon_t$
    \Else
      \State \texttt{new} $\gets \texttt{prev}$
    \EndIf

    \State \texttt{reward}$[t,a] \gets \texttt{new}$
  \EndFor
\EndFor
\end{algorithmic}

\caption{Simulation algorithm}
\label{alg:agent_simulation}
\end{minipage}
\end{figure}

\section{Sensitivity to cohort size and cohort threshold limits in the case of social comparison}
\label{appendix:robustness}

\begin{figure}[h!]
\centering
\includegraphics[width=\linewidth]{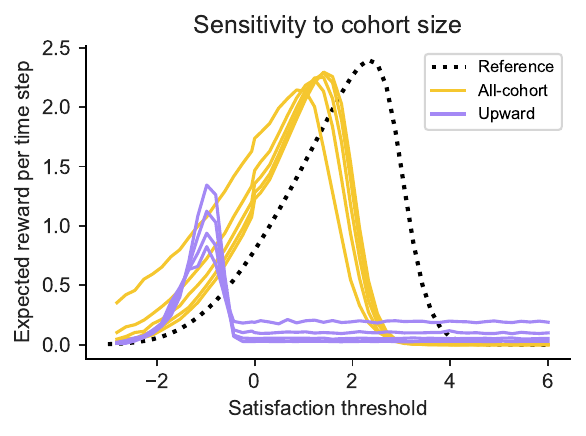}
\caption{ Social comparison penalizes high satisfaction thresholds, regardless of cohort size. The figure shows the mean rewards for all-cohort comparison and upward social comparison on a rough, non-skewed landscape, for cohort sizes of 10, 25, 50, 100, and 200 agents. $\epsilon_t \sim \mathcal{N}(\mu,\sigma^2)$, $\varphi=0$, $t_{\text{max}}=1000$, cohorts of $100$ agents, with thresholds uniformly sampled from $[-3,6]$, averaged over $10^3$ simulations. }
\label{fig:cohort-size}
\end{figure}

\begin{figure}[b!]
\centering
\includegraphics[width=\linewidth]{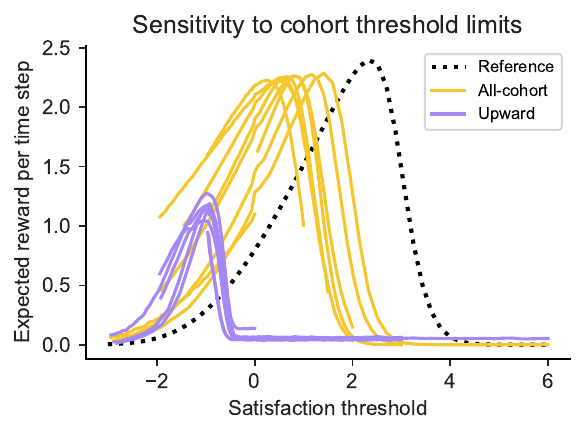}
\caption{Social comparison penalizes high satisfaction thresholds, regardless of threshold limits. The figure shows the mean rewards for all-cohort comparison and upward social comparison on a rough, non-skewed landscape, for a cohort size of 100 and with threshold limits samples uniformly on the intervals: $[-3,6]$, $[-3,4]$, $[-3,0]$, $[-2,1]$, $[-1.5,1.5]$, and $[-1,2]$. $\epsilon_t \sim \mathcal{N}(\mu,\sigma^2)$, $\varphi=0$, $t_{\text{max}}=1000$, cohorts of $100$ agents, with thresholds uniformly sampled from $[-3,6]$, averaged over $10^4$ simulations.}
\label{fig:threshold-limits}
\end{figure}

The qualitative behavior of the model in the case of social comparison is not sensitive to cohort sizes and cohort threshold limits. Fig.~\ref{fig:cohort-size} shows the social comparison behavior for different cohort sizes. Fig.~\ref{fig:threshold-limits} shows the social comparison behavior for different threshold limit intervals. Regardless of cohort size and threshold interval, all-cohort comparison leads to higher average rewards and higher optimal thresholds than upward social comparison. Both modes of social comparison exhibit lower average reward and lower optimal threshold than the idealized reference landscape where the agents have perfect information about the mean and variance.

\clearpage

\bibliography{main}

\end{document}